\documentclass[12pt]{revtex4}
\usepackage{amsfonts}
\usepackage{graphicx, amsmath}
\usepackage{longtable}
\makeatletter

\newcommand{\Rmnum}[1]{\expandafter\@slowromancap\romannumeral #1@}
\makeatother
\begin{document}
\title[Short Title]{Transitionless-based shortcuts for rapidly generating two-atom 3D entanglement}
\author{Rui Peng}
\author{Yue Zheng}
\author{Si-Wen Liu}
\author{Xiao-Pan Li}
\author{Jin-Lei Wu}
\author{Xin Ji\footnote{E-mail: jixin@ybu.edu.cn}}
\affiliation{Department of Physics, College of Science, Yanbian University, Yanji, Jilin 133002, People's Republic of China}

\begin{abstract}
\noindent \textbf{{Abstract}} An experimentally feasible scheme is proposed for rapidly generating two-atom three-dimensional ~(3D) entanglement with one step. As one technique of shortcuts to adiabaticity, transitionless quantum driving is applied to speed up the adiabatic generation of two-atom 3D entanglement. Apart from the rapid rate, the scheme has much higher experimental feasibility than the recent research (Quant. Inf. Process. DOI: 10.1007/s11128-016-1453-2, 2016). Besides, numerical simulations indicate the scheme has strong robustness against parameter deviations and decoherence.
\\{\bf{Keywords:}} 3D entanglement, Shortcuts to adiabaticity, Transitionless quantum driving
\end{abstract}
\maketitle
\section{Introduction}
Recently, because high-dimensional entanglement has more superior security than qubit entanglement in the aspect of quantum key distribution~\cite{BKB2001,BM2002,CBK2002,YW2016}, researchers have paid more and more attentions to investigations of high-dimensional entanglement. Besides, entangled high-dimensional systems are stronger than two-dimensional systems in violations of local realism~\cite{DPMW2000}. Hence, the generation of high-dimensional entanglement is increasingly concerned by researchers via various techniques, such as stimulated Raman adiabatic passage~(STIRAP)~\cite{LP2012,YLS2015,SSW2016}, quantum Zeno dynamics~\cite{WG2011,SJR2013,QCW2013}, and dissipative dynamics~\cite{XQ201402,SL2014}. Among these techniques, STIRAP is popular in fields of time-dependent interaction for robust quantum state transfer~\cite{KHB1998,PIM2007}. In recent several years, various entanglement generations have been come up with based on STIRAP. For example, a generation of a three-atom singlet state was implemented by Lu $et~al.$ in 2013~\cite{MYJ2013}; a NOON state was created by Liu $et~al.$ in 2014~\cite{QQC2014}; the preparation of multi-qubit W states was achieved by Wei $et~al.$ in 2015~\cite{XM2015}; $n$-qubit GHZ states were implemented in 2016~\cite{JCJ2016}.

As we all know, however, STIRAP technique usually requires a relatively long interaction time which may accumulate decoherence factors and errors leading to invalid dynamics. Many efforts have been made to accelerate an adiabatic quantum evolution process, and thus a set of techniques called ``shortcuts to adiabaticity~(STA)'' arise at the historic moment~\cite{XASA2010,AC2013,XCE2011,XJ2012,SEX2014,TE2013,SLX2015,XHQ2016,YHC2016,AHA2016}. By using these techniques, lots of remarkable achievements have been made in quantum information processing, such as fast quantum state transfer~\cite{YHC2014,YYQ2014,JJS2016}, fast entanglement generations~\cite{YYJ2015,CSC2016,LXX2016,WYY2016} and fast quantum gate constructions~\cite{YLQ2015,YLC2015,YLX2015,JTD2015,YYQ2015}. Also, many generations of high-dimensional entanglement have been achieved~\cite{JYC2016,ZYY2016,WSJ2016,HSW2016,WJZ2016}, in which using transitionless quantum driving~(TQD), Chen $et~al.$ generated a three-atom singlet state~\cite{ZYY2016} and He $et~al.$ generated a two-atom 3D entangled state~\cite{HSW2016}; using Lewis-Riesenfeld invariants~(LRI), Lin $et~al.$~\cite{JYC2016} and we~\cite{WSJ2016} generated two-atom 3D entangled states, respectively; using both of TQD and LRI, we generated three-atom tree-type 3D entangled states~\cite{WJZ2016}.

Very recently, Yang $et~al.$ proposed a shortcut scheme for rapid generation of a two-atom 3D entangled state by using LRI technique~\cite{YFY2016}. The scheme overcomes shortcomings of the atom-cavity-fiber system in Refs.~\cite{YLS2015,JYC2016} and rapidly generated 3D entangled state with only one step, which may enhance experimental feasibility. However, Ref.~\cite{YFY2016} ignores the experimental feasibility of driving pulses. Even though the functions of driving pulses in Ref.~\cite{YFY2016} are superpositions of simple Gaussian functions or sinusoidal functions by curve fitting, driving pulses used in the scheme are only truncations of corresponding pulse superpositions. In other words, the pulses applied in Ref.~\cite{YFY2016} do not include a complete Gaussian or sinusoidal form and are not smoothly turned on or off, which will be a great challenge in experiment. Therefore, it is essential to optimize driving pulses for improvement of experimental feasibility. Fortunately, we find great illumination in Ref.~\cite{MYL2014} which uses TQD to speed up two-atom population transfer and creation of maximum entanglement. In this work, we put TQD into rapid generation of a two-atom 3D entangled state, and obtain the target state with very high fidelity and quite short runtime. More importantly, driving pulses in the TQD scheme are superpositions of complete Gaussian pulses, which ensures much more feasibility in practice.
\section{Physical model and effective dynamics}\label{a}

\begin{figure}[htb]\centering
  \includegraphics[width=10cm]{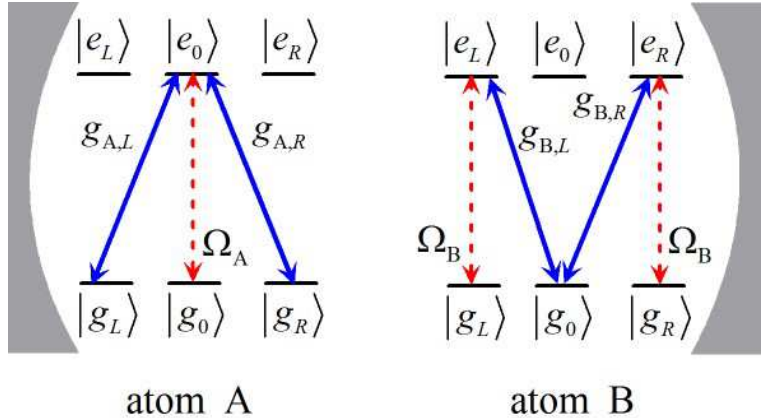}\\
  \caption{The diagrammatic sketch of atom-cavity system, atomic level configurations and related transitions}\label{f1}
\end{figure}
As shown in Fig.~\ref{f1}, there are two atoms trapped in a bimodule optical cavity. Atom \rm A has one upper level $|e_0\rangle$ and three lower levels $|g_L\rangle$, $|g_0\rangle$ and $|g_R\rangle$. Atom \rm B has the same lower levels as atom \rm A and two upper levels $|e_L\rangle$ and $|e_R\rangle$. For atom \rm A, the transition $|e_0\rangle_{\rm A}\leftrightarrow|g_{{L(R)}}\rangle_{\rm A}$ is resonantly coupled to the left-circularly~(right-circularly) polarized cavity mode with coupling constant $g_{{\rm A}, L(R)}$, and $|e_0\rangle_{\rm A}\leftrightarrow|g_0\rangle_{\rm A}$ is resonantly driven by a classical time-dependent laser field with the Rabi frequency $\Omega_{\rm A}(t)$. For atom \rm B, $|e_{L(R)}\rangle_{\rm B}\leftrightarrow|g_0\rangle_{\rm B}$ is resonantly coupled to the left-circularly~(right-circularly) polarized cavity mode with coupling constant $g_{{\rm B}, L(R)}$, and $|e_{L(R)}\rangle_{\rm B}\leftrightarrow|g_{L(R)}\rangle_{\rm B}$ is resonantly driven by a classical time-dependent laser field with the Rabi frequency $\Omega_{\rm B}(t)$.
Then, the interaction Hamiltonian of the atom-cavity system is~($\hbar=1$):
\begin{eqnarray}\label{e1}
H(t)=\Omega_{\rm A}(t)|g_0\rangle_{\rm A}\langle{e_0}|+\sum_{i=L,R}\left[\Omega_{\rm B}(t)|g_i\rangle_{\rm B}\langle{e_i}|+g_{{\rm A},i}a_{i}|e_0\rangle_{\rm A}\langle{g_i}|+g_{{\rm B},i}a_{i}|e_i\rangle_{\rm B}\langle{g_0}|\right]
+\rm H.c.,
\end{eqnarray}
where $a_{L(R)}$ is the annihilation operator of left-circularly~(right-circularly) polarized cavity mode. We assume $g_L=g_R=g$ and $g$ is real for simplicity.

The system is initially in the state $|\phi_1\rangle=|g_0\rangle_{\rm A}|g_0\rangle_{\rm B}|0\rangle_c$ which denotes two atoms in the states $|g_0\rangle_{\rm A}$ and $|g_0\rangle_{\rm B}$, respectively, and the cavity in the vacuum state. The atom-cavity system will evolve in the subspace spanned by
\begin{eqnarray}\label{e2}
|\phi_1\rangle&=&|g_0\rangle_{\rm A}|g_0\rangle_{\rm B}|0\rangle_c,\quad
|\phi_2\rangle=|e_0\rangle_{\rm A}|g_0\rangle_{\rm B}|0\rangle_c,\quad
|\phi_3\rangle=|g_L\rangle_{\rm A}|g_0\rangle_{\rm B}|L\rangle_c,\nonumber\\
|\phi_4\rangle&=&|g_R\rangle_{\rm A}|g_0\rangle_{\rm B}|R\rangle_c,\quad
|\phi_5\rangle=|g_L\rangle_{\rm A}|e_L\rangle_{\rm B}|0\rangle_c,\quad
|\phi_{6}\rangle=|g_R\rangle_{\rm A}|e_R\rangle_{\rm B}|0\rangle_c,\nonumber\\
|\phi_{7}\rangle&=&|g_L\rangle_{\rm A}|g_L\rangle_{\rm B}|0\rangle_c,\quad
|\phi_{8}\rangle=|g_R\rangle_{\rm A}|g_R\rangle_{\rm B}|0\rangle_c.
\end{eqnarray}
in which $|L(R)\rangle_{c}$ denotes a single left-circularly~(right-circularly) polarized photon in the cavity. Hamiltonian (\ref{e1}) can be rewritten by
\begin{eqnarray}\label{e3}
H(t)&=&\Omega_{\rm A}(t)|\phi_1\rangle\langle\phi_2|+\Omega_{\rm B}(t)(|\phi_5\rangle\langle\phi_7|+|\phi_6\rangle\langle\phi_8|)\nonumber\\
&&+g(|\phi_2\rangle\langle\phi_3|+|\phi_2\rangle\langle\phi_4|+|\phi_3\rangle\langle\phi_5|+|\phi_4\rangle\langle\phi_6|)+\rm H.c.,
\end{eqnarray}
If we set
\begin{eqnarray}\label{e4}
|\psi_1\rangle=\frac{1}{\sqrt2}(|\phi_3\rangle+|\phi_4\rangle),
\quad|\psi_2\rangle=\frac{1}{\sqrt2}(|\phi_5\rangle+|\phi_6\rangle),
\quad|\psi_3\rangle=\frac{1}{\sqrt2}(|\phi_7\rangle+|\phi_8\rangle),\nonumber\\
|\psi_1^-\rangle=\frac{1}{\sqrt2}(|\phi_3\rangle-|\phi_4\rangle),
\quad|\psi_2^-\rangle=\frac{1}{\sqrt2}(|\phi_5\rangle-|\phi_6\rangle),
\quad|\psi_3^-\rangle=\frac{1}{\sqrt2}(|\phi_7\rangle-|\phi_8\rangle),
\end{eqnarray}
Hamiltonian (\ref{e3}) will become
\begin{eqnarray}\label{e5}
H(t)&=&\Omega_{\rm A}(t)|\phi_1\rangle\langle\phi_2|+\Omega_{\rm B}(t)|\psi_2\rangle\langle\psi_3|+\Omega_{\rm B}(t)|\psi_2^-\rangle\langle\psi_3^-|\nonumber\\
&&+\sqrt2g|\phi_2\rangle\langle\psi_1|+g|\psi_1\rangle\langle\psi_2|+g|\psi_1^-\rangle\langle\psi_2^-|+\rm H.c..
\end{eqnarray}
When the initial state is $|\phi_1\rangle$, the system evolution does not involve $|\psi_1^-\rangle$, $|\psi_2^-\rangle$ and $|\psi_3^-\rangle$, so Hamiltonian (\ref{e5}) becomes
\begin{eqnarray}\label{e6}
H(t)&=&\Omega_{\rm A}(t)|\phi_1\rangle\langle\phi_2|+\sqrt2g|\phi_2\rangle\langle\psi_1|+g|\psi_1\rangle\langle\psi_2|+\Omega_{\rm B}(t)|\psi_2\rangle\langle\psi_3|+\rm H.c..
\end{eqnarray}
Next, when introducing the following transformations
\begin{eqnarray}\label{e7}
|\Psi_d\rangle=\frac{1}{\sqrt3}(|\phi_2\rangle-\sqrt2|\psi_2\rangle),
\quad|\Psi_\pm\rangle=\frac{1}{\sqrt6}(\sqrt2|\phi_2\rangle\pm\sqrt3|\psi_1\rangle+|\psi_2\rangle),
\end{eqnarray}
Hamiltonian (\ref{e6}) has the form
\begin{eqnarray}\label{e8}
&H(t)=H_0+V(t),\nonumber\\
&H_0=\sqrt3g(|\Psi_+\rangle\langle\Psi_+|-|\Psi_-\rangle\langle\Psi_-|)\nonumber\\
&V(t)=\frac{\Omega_{\rm A}(t)}{\sqrt3}|\phi_1\rangle\left(\langle\Psi_d|+\langle\Psi_+|+\langle\Psi_-|\right)+\frac{\Omega_{\rm B}(t)}{\sqrt3}\left[-\sqrt2|\Psi_d\rangle+\frac{1}{\sqrt2}\left(|\Psi_+\rangle+|\Psi_-\rangle\right)\right]\langle\psi_3|+\rm H.c..\nonumber\\
\end{eqnarray}
Trough performing the unitary transformation $U=\exp(-iH_0t)$, we obtain the interaction Hamiltonian with respect to $H_0$ as below
\begin{eqnarray}\label{e9}
H_I(t)&=&\frac{\Omega_{\rm A}(t)}{\sqrt3}|\phi_1\rangle\left(\langle\Psi_d|+\langle\Psi_+|e^{-i\sqrt3gt}+\langle\Psi_-|e^{i\sqrt3gt}\right)\nonumber\\
&&+\frac{\Omega_{\rm B}(t)}{\sqrt3}\left[-\sqrt2|\Psi_d\rangle+\frac{1}{\sqrt2}\left(|\Psi_+\rangle e^{i\sqrt3gt}+|\Psi_-\rangle e^{-i\sqrt3gt}\right)\right]\langle\psi_3|+\rm H.c..
\end{eqnarray}
Then by neglecting the terms with high oscillating frequency under the condition $\Omega_{\rm A}(t),\Omega_{\rm B}(t)/\sqrt2\ll3g$, we obtain an effective Hamiltonian
\begin{eqnarray}\label{e10}
H_e(t)=\frac{1}{\sqrt3}\Omega_{\rm A}(t)|\phi_1\rangle\langle\Psi_d|-\frac{\sqrt2}{\sqrt3}\Omega_{\rm B}(t)|\Psi_d\rangle\langle\psi_3|+\rm H.c..
\end{eqnarray}
Instantaneous eigenstates of Hamiltonian (\ref{e10}) corresponding to the eigenvalues $\lambda_0=0$ and $\lambda_{\pm}=\pm \Omega(t)/\sqrt{3}$, respectively, are
\begin{eqnarray}\label{e11}
|n_0(t)\rangle=\left(
\begin{array}{c}
-\cos\theta(t)\\
0\\
\sin\theta(t)
\end{array}\right),\quad
|n_\pm(t)\rangle=\frac{1}{\sqrt{2}}\left(
\begin{array}{c}
\sin\theta(t)\\
\pm 1\\
\cos\theta(t)
\end{array}\right),
\end{eqnarray}
with $\Omega(t)=\sqrt{\Omega_{\rm A}(t)^2+2\Omega_{\rm B}(t)^2}$ and $\tan\theta(t)=-\Omega_{\rm A}(t)/\sqrt2\Omega_{\rm B}(t)$. With STIRAP technique, if the adiabatic criterion $|\dot{\theta}(t)|\ll\sqrt{2}\Omega(t)$ and boundary conditions $\theta(0)=0$ and $\theta(t_f)=-\arctan\sqrt2$~($t_f$ is the final time) are satisfied well, the effective system evolution will follow the dark state $|n_0(t)\rangle$ and we can obtain the target state $|\Psi_{3D}\rangle=\frac{1}{\sqrt3}(|\phi_1\rangle+|\phi_7\rangle+|\phi_8\rangle)$ through the desired evolution $|\phi_1\rangle\rightarrow|\Psi_{3D}\rangle=\frac{1}{\sqrt3}(|\phi_1\rangle+\sqrt2|\psi_3\rangle)$.
\section{TQD scheme for rapidly generating two-atom 3D entanglement}\label{b}
\begin{figure}[htb]\centering
\centering
\includegraphics[width=10cm]{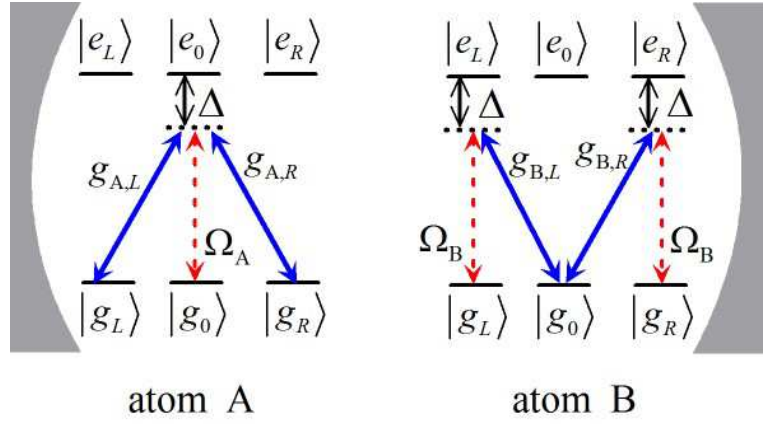}
\caption{The diagrammatic sketch of the alternative system for TQD scheme}\label{f2}
\end{figure}
When the operation time is drastically shortened, the adiabatic criterion $|\dot{\theta}(t)|\ll\sqrt{2}\Omega(t)$ will be destroyed and the system evolution does not follow one of instantaneous eigenstates $|n_k(t)\rangle~(k=0,\pm)$ any more. In other words, instantaneous eigenstates do not satisfy the Schr\"{o}dinger equation $i\partial_t|n_k(t)\rangle=H_e(t)|n_k(t)\rangle$. By using Berry's transitionless tracking algorithm~\cite{Berry2009}, we can reverse engineer a transitionless Hamiltonian $H_{TQD}(t)$, which can drive the system evolution to follow one of the eigenstates exactly. From Ref.~\cite{Berry2009}, the
simplest form of $H_{TQD}(t)$ is
\begin{eqnarray}\label{e12}
H_{TQD}(t)=i\sum_{k=0,\pm}|\partial_ {t}n_k(t)\rangle\langle n_k(t)|.
\end{eqnarray}
Through substituting Eq.~(\ref{e11}) into Eq.~(\ref{e12}), we obtain
\begin{eqnarray}\label{e13}
H_{TQD}(t)=i\dot{\theta}(t)|\phi_1\rangle\langle \psi_3|+\rm H.c.,
\end{eqnarray}
with $\dot{\theta}(t)=\sqrt2[\Omega_{\rm A}(t)\dot\Omega_{\rm B}(t)-\dot\Omega_{\rm A}(t)\Omega_{\rm B}(t)]/\Omega(t)^2$. However, the system we investigate now is so complicated that such Hamiltonian~(\ref{e13}) is impossibly achieved. Therefore, we must find an alternative equivalent system. The most common way is to change all resonant atomic transitions to non-resonant transitions with a large detuning parameter $\Delta$~\cite{ZYY2016,HSW2016,WJZ2016,MYL2014}. The diagrammatic sketch of the alternative system is shown in Fig.~\ref{f2}.

Correspondingly, the interaction Hamiltonian of the alternative system is
\begin{eqnarray}\label{e14}
H'(t)&=&\{\Omega'_{\rm A}(t)|g_0\rangle_{\rm A}\langle{e_0}|+\sum_{i=L,R}\left[\Omega'_{\rm B}(t)|g_i\rangle_{\rm B}\langle{e_i}|+g_{{\rm A},i}a_{i}|e_0\rangle_{\rm A}\langle{g_i}|+g_{{\rm B},i}a_{i}|e_i\rangle_{\rm B}\langle{g_0}|\right]
+\rm H.c.\}\nonumber\\&&+\Delta(|e_0\rangle_{\rm A}\langle{e_0}|+|e_L\rangle_{\rm B}\langle{e_L}|+|e_R\rangle_{\rm B}\langle{e_R}|).
\end{eqnarray}
Then analogous to the process from Eq.~(\ref{e1}) to Eq.~(\ref{e10}), the effective Hamiltonian for the alternative system is written as
\begin{eqnarray}\label{e15}
H'_e(t)=[\frac{1}{\sqrt3}\Omega'_{\rm A}(t)|\phi_1\rangle\langle\Psi_d|-\frac{\sqrt2}{\sqrt3}\Omega'_{\rm B}(t)|\Psi_d\rangle\langle\psi_3|+\rm H.c.]+\Delta|\Psi_d\rangle\langle{\Psi_d}|.
\end{eqnarray}
By introducing the large detuning condition $|\Delta|\gg|\Omega'_{\rm A}(t)/\sqrt3|,|\sqrt2\Omega'_{\rm B}(t)/\sqrt3|$ and adiabatically eliminating $|\Psi_d\rangle$, the effective Hamiltonian (\ref{e15}) is simplified to a new effective Hamiltonian
\begin{eqnarray}\label{e16}
H_{eff}(t)=\frac{|\Omega'_{\rm A}(t)|^2}{3\Delta}|\phi_1\rangle\langle\phi_1|+\frac{2|\Omega'_{\rm B}(t)|^2}{3\Delta}|\psi_3\rangle\langle\psi_3|+[\frac{\sqrt2\Omega'_{\rm A}(t)\Omega'_{\rm B}(t)^\ast}{3\Delta}|\phi_1\rangle\langle\psi_3|+\rm H.c.],
\end{eqnarray}
in which the superscript $^\ast$ denotes the complex conjugate. The first two terms can be removed by setting $\Omega'_{\rm B}(t)=i\Omega'_{\rm A}(t)/\sqrt2$ and the effective Hamiltonian (\ref{e16}) becomes
\begin{eqnarray}\label{e17}
H'_{eff}(t)=i\frac{\Omega'_{\rm A}(t)^2}{3\Delta}|\phi_1\rangle\langle\psi_3|+\rm H.c..
\end{eqnarray}
The effective Hamiltonian $H'_{eff}(t)$ is exactly equivalent to the transitionless Hamiltonian (\ref{e13}) when $\dot{\theta}(t)=\Omega'_{\rm A}(t)^2/3\Delta$, i.e.,
\begin{eqnarray}\label{e18}
\Omega'_{\rm A}(t)=\sqrt{3\sqrt2\Delta[\Omega_{\rm A}(t)\dot\Omega_{\rm B}(t)-\dot\Omega_{\rm A}(t)\Omega_{\rm B}(t)]/\Omega(t)^2},
\end{eqnarray}
which is the correlation between TQD and STIRAP with respect to Rabi frequencies. Hamiltonian (\ref{e17}) can drive the system evolution to follow one of the instantaneous eigenstates $|n_k(t)\rangle$ even without the adiabatic criterion. Therefore, as long as the boundary conditions $\theta(0)=0$ and $\theta(t_f)=-\arctan\sqrt2$ are satisfied well, the target state $|\Psi_{3D}\rangle$ can be achieved.
\section{Numerical simulations and discussions}\label{c}
\begin{figure}[htb]\centering
  \includegraphics[width=\linewidth]{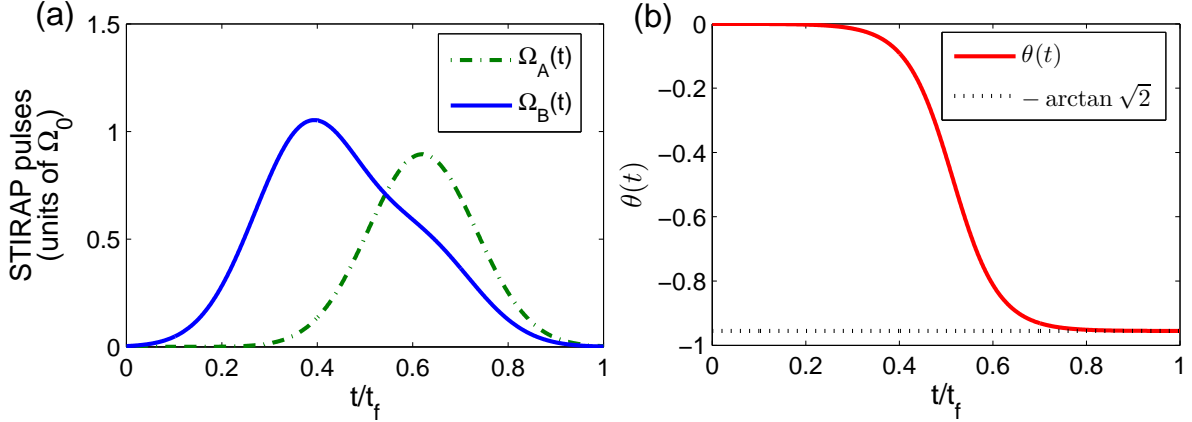}\\
  \caption{{\bf a}~Time dependence of STIRAP pulses. {\bf b}~Time dependence of $\theta(t)$}\label{f3}
\end{figure}
First of all, in order to meet the boundary conditions, STIRAP Rabi frequencies $\Omega_{\rm A}(t)$ and $\Omega_{\rm B}(t)$ can be chosen as a Gaussian pulse and superposition of two Gaussian pulses, respectively~\cite{NTB2001}
\begin{eqnarray}\label{e19}
\Omega_{\rm A}(t)&=&\frac{2}{\sqrt5}\Omega_0e^{-(t-t_f/2-\tau)^2/T^2},\nonumber\\
\Omega_{\rm B}(t)&=&\frac{1}{\sqrt5}\Omega_0e^{-(t-t_f/2-\tau)^2/T^2}+\Omega_0e^{-(t-t_f/2+\tau)^2/T^2},
\end{eqnarray}
where $\Omega_0$ is the amplitude of Gaussian pulses and $\tau=0.12t_f$ and $T=0.16t_f$ are two related parameters of Gaussian pulses. $\Omega_{\rm A}(t)$ and $\Omega_{\rm B}(t)$ are shown in Fig.~\ref{f3}a, and corresponding $\theta(t)$ is given in Fig.~\ref{f3}b, which clearly shows that the boundary conditions $\theta(0)=0$ and $\theta(t_f)=-\arctan\sqrt2$ are satisfied perfectly.

\begin{figure}[htb]\centering
\includegraphics[width=\linewidth]{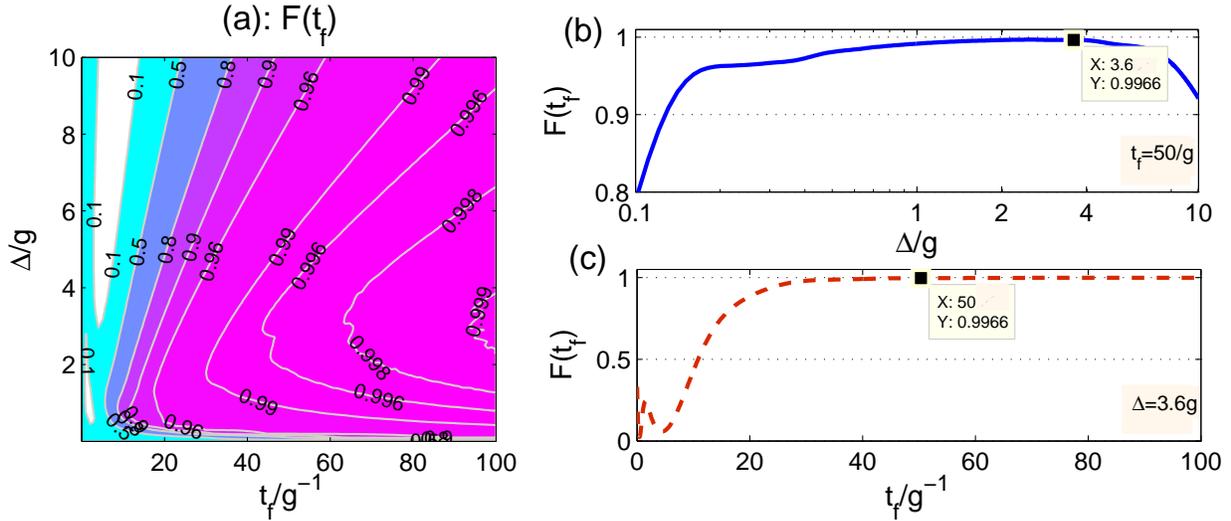}\\
\caption{Final fidelity versus {\bf a}~$t_f/g^{-1}$ and $\Delta/g$; {\bf b}~$\Delta/g$ with $t_f=50/g$; {\bf c}~$t_f/g^{-1}$ with $\Delta=3.6g$}\label{f4}
\end{figure}
For meeting the limit condition $\Omega_{\rm A}(t),\Omega_{\rm B}(t)/\sqrt2\ll3g$, we choose the same value $\Omega_0=0.35g$ as in Ref.~\cite{YFY2016}. Then in order to choose suitable values of $t_f$ and $\Delta$, in Fig.~\ref{f4}a, we plot a contour image of the final fidelity versus $t_f/g^{-1}$ and $\Delta/g$. The final fidelity is defined by $F(t_f)=|\langle\Psi_{\rm 3D}|\Psi(t_f)\rangle|^2$, for which $|\Psi(t_f)\rangle$ is the state of the system at $t=t_f$ governed by Hamiltonian (\ref{e14}) for TQD scheme. Form Fig.~\ref{f4}a, we learn that there is a wide range of values of $t_f$ and $\Delta$ ensuring a high final fidelity. More clearly, in Fig.~\ref{f4}b, we plot the relationship between $F(t_f)$ and $\Delta$ with $t_f=50/g$ which is the same as in Ref.~\cite{YFY2016}. In Fig.~\ref{f4}c, we plot $F(t_f)$ versus $t_f$ with $\Delta=3.6g$ which can guarantee a quite high $F(t_f)$. From Fig.~\ref{f4}c, we know that one can adopt a pair of parameters $\Delta=3.6g$ and $t_f=50/g$ for the following discussions.

\begin{figure}[htb]\centering
  \includegraphics[width=\linewidth]{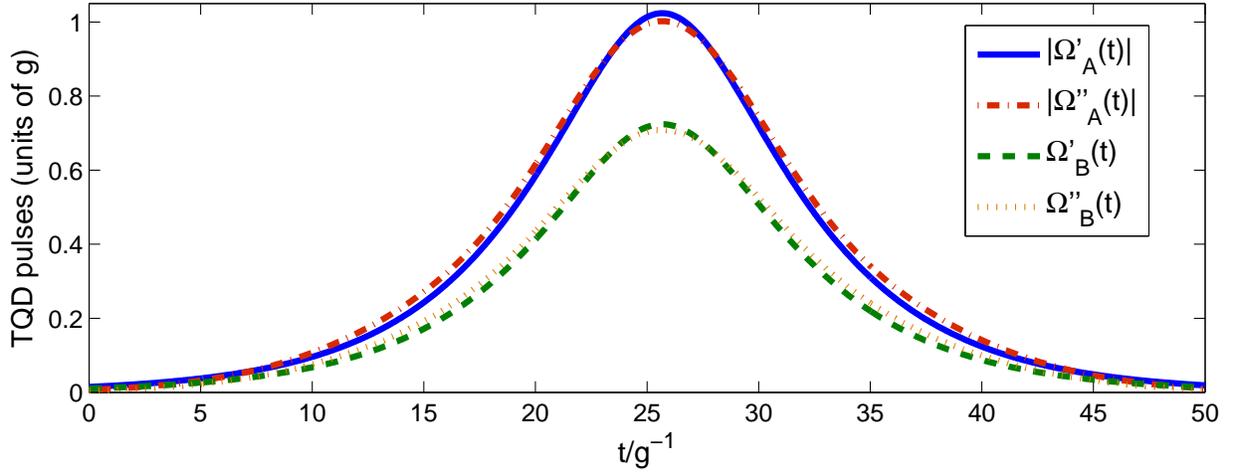}\\
  \caption{Time dependence of TQD pulses}\label{f5}
\end{figure}
Next, we test the feasibility of driving pulses of TQD scheme by plotting the time dependence of $\Omega'_{\rm A}(t)$ and $\Omega'_{\rm B}(t)$ in Fig.~\ref{f5}. From Fig.~\ref{f5}, we can see that $\Omega'_{\rm A}(t)$ and $\Omega'_{\rm B}(t)$ are both turned on and off near smoothly. Furthermore, since the function of $\Omega'_{\rm A}(t)$ or $\Omega'_{\rm B}(t)$ is so complicated that it is difficultly obtained in experiment, we replace $\Omega'_{\rm B}(t)$ with a superimposition of two Gaussian pulses by curves fitting. The alternative pulse of $\Omega'_{\rm B}(t)$ is
\begin{eqnarray}\label{e20}
\Omega''_{\rm B}(t)&=&\Omega_{1}e^{-(t-\tau_{1})^2/\chi_{1}^2}+\Omega_{2}e^{-(t-\tau_{2})^2/\chi_{2}^2},
\end{eqnarray}
with related parameters
\begin{eqnarray}\label{e21}
\Omega_{1}=0.3861g,\quad\Omega_{2}=0.3227g,\quad\tau_{1}=25.6816/g,\nonumber\\
\tau_{2}=25.6808/g,\quad\chi_{1}=12.2827/g,\quad\chi_{2}=5.7835/g.
\end{eqnarray}
The alternative pulse of $\Omega'_{\rm A}(t)$ is $\Omega''_{\rm A}(t)=-i\sqrt2\Omega''_{\rm B}(t)$. The time dependence of $\Omega''_{\rm A}(t)$ and $\Omega''_{\rm B}(t)$ is also shown in Fig.~\ref{f5}, and highly approximate coincidences of two pair of curves prove that the alternative pulses are pretty effective. In addition, it is worth mentioning that the function either $\Omega_{1}e^{-(t-\tau_{1})^2/\chi_{1}^2}$ or $\Omega_{2}e^{-(t-\tau_{2})^2/\chi_{2}^2}$ has a near complete Gaussian curve from $t=0$ to $t=t_f$, which guarantees the high feasibility in experiment.

\begin{figure}[htb]\centering
  \includegraphics[width=\linewidth]{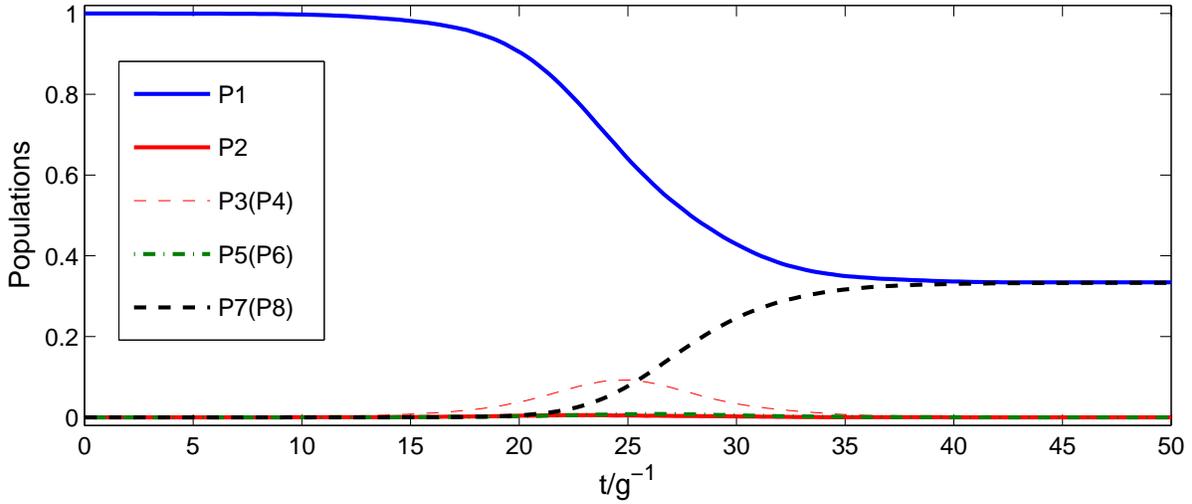}\\
  \caption{Time evolutions of the populations of the states in Eq.~\ref{e2}}\label{f6}
\end{figure}

\begin{figure}[htb]\centering
  \includegraphics[width=\linewidth]{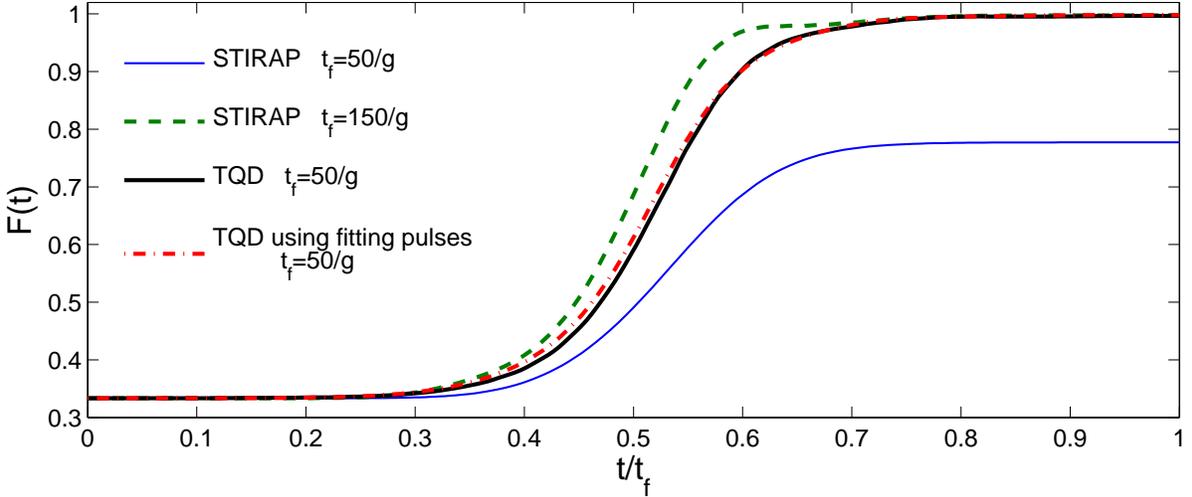}\\
  \caption{Fidelity for the target state $|\Psi_{\rm 3D}\rangle$ with different methods versus $t/t_f$}\label{f7}
\end{figure}

To further test the availability of two alternative pulses, in Fig.~\ref{f6}, we plot time evolutions of the population of the states in Eq.~(\ref{e2}). As shown in Fig.~\ref{f6}, the target state $|\Psi_{\rm 3D}\rangle$ is exactly obtained at $t=t_f$, but other states are slightly populated.
To show that the TQD scheme is faster than STIRAP, we plot the fidelity for the target state $|\Psi_{\rm 3D}\rangle$ with
different methods versus $t/t_f$ in Fig.~\ref{f7}. As we can see from Fig.~\ref{f7}, STIRAP does require much longer interaction time than TQD for implementing the target state $|\Psi_{\rm 3D}\rangle$. Figure~\ref{f7} also proves that the fitting pulses are pretty effective.

\begin{figure}[htb]\centering
  \includegraphics[width=\linewidth]{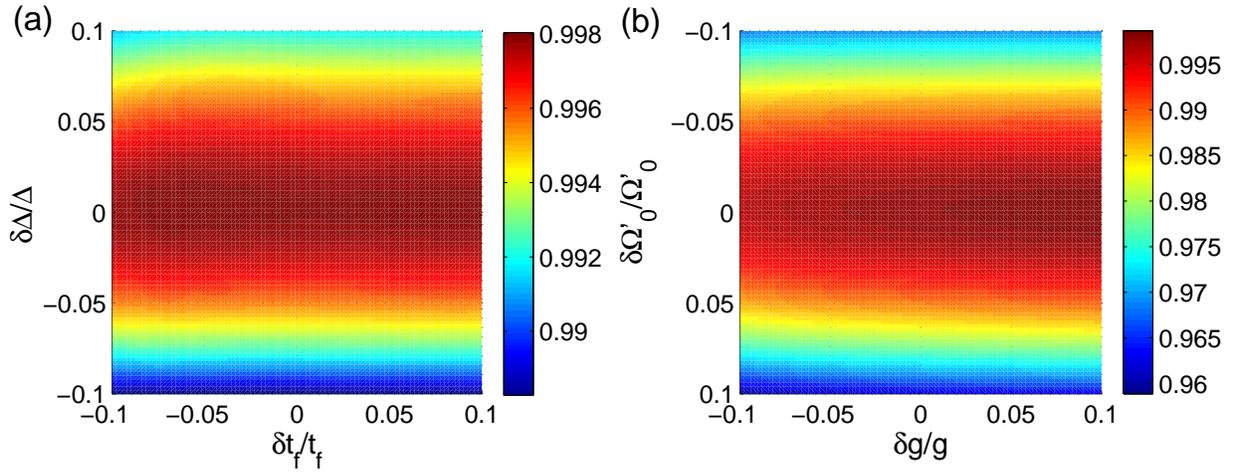}\\
  \caption{Effects of the variations in the various parameters on the final fidelity.}\label{f8}
\end{figure}

In the discussions above, the controlling of parameters is perfect and the system is considered as isolated from environment, which is impossible in a real experiment. Therefore, we are supposed to give discussions about the robustness of the TQD scheme against parameter errors and decoherence caused by atomic spontaneous emissions and the cavity photon leakage. Here we define $\delta x=x'-x$ as the deviation of $x$, for which $x$ denotes the ideal value and $x'$ denotes the actual value. In Fig.~\ref{f8}, we consider the effects of the variations in the parameters on the final fidelity for generating the target state $|\Psi_{\rm 3D}\rangle$, in which the variation in $\Omega'_0$ denotes the collective variations in $\Omega_{1}$ and $\Omega_{2}$. As shown in Fig.~\ref{f8}, the TQD scheme for generating 3D entanglement $|\Psi_{\rm 3D}\rangle$ is robust against the variations in control parameters. Besides, Figs.~\ref{f8}a and \ref{f8}b also show that the TQD scheme is hardly effected by the variation in $t_f$ and $g$.

\begin{figure}[htb]\centering
  \includegraphics[width=\linewidth]{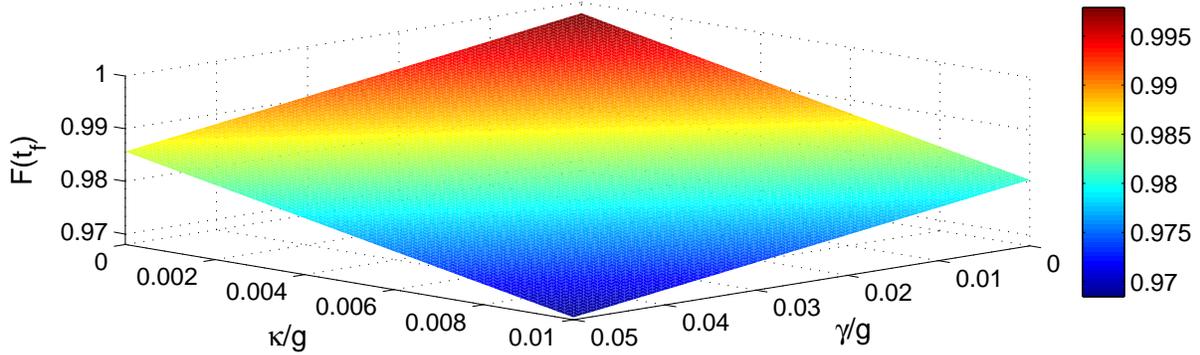}\\
  \caption{Final fidelity versus $\kappa/g$ and $\gamma/g$.}\label{f9}
\end{figure}
Then taking decoherence caused by atomic spontaneous emissions and the cavity photon leakage into account, the whole system is dominated by the master equation
\begin{eqnarray}\label{e22}
\dot{\rho}(t)&=&-i[H'(t),\rho(t)]\nonumber\\
&&-\sum_{j=L,R}\frac{\kappa_{j}}{2}[a_{j}^{\dag}a_{j}\rho(t)-2a_{j}\rho(t)a_{j}^{\dag}+\rho(t)a_{j}^\dag
a_{ij} ]\nonumber\\
&&-\sum_{i=0,L,R}\frac{\gamma_{i}^{\rm A}}{2}[\sigma^{\rm A}_{e_{0}}\rho(t)-2\sigma^{\rm A}_{g_i,e_0}\rho(t)\sigma^{\rm A}_{e_0,g_i}+\rho(t)\sigma^{\rm A}_{e_0,e_0}]
\nonumber\\
&&-\sum_{i=0,L,R}\sum_{j=L,R}\frac{\gamma_{i,j}^{\rm B}}{2}[\sigma^{\rm B}_{e_j,e_j}\rho(t)-2\sigma^{\rm B}_{g_i,e_j}\rho(t)\sigma^{\rm B}_{e_j,g_i}+\rho(t)\sigma^{\rm B}_{e_j,e_j}],
\end{eqnarray}
where $\kappa_{j}$ is the cavity photon leakage rate of $j$-circular polarization mode; $\gamma_{i}^{\rm A}$ is the spontaneous emission rate of atom \rm A from the excited state $|e_0\rangle$ to the ground state $|g_j\rangle$; $\gamma_{i,j}^{\rm B}$ is the spontaneous emission rate of atom \rm B from the excited state $|e_j\rangle$ to the ground state $|g_i\rangle$; $\sigma_{e_j,e_j}=|e_j\rangle\langle e_j|$, and $\sigma_{e_j,g_i}=|e_j\rangle\langle g_i|$. For simplicity, we assume $\kappa_{L}=\kappa_{R}=\kappa$ and $\gamma_{i}^{\rm A}=\gamma_{i,j}^{\rm B}=\gamma/2$. In Fig.~\ref{f9}, we plot the final fidelity for generating two-atom 3D entanglement versus $\kappa/g$ and $\gamma/g$. As we can see from Fig.~\ref{f9}, we know that atomic spontaneous emissions spoil the TQD scheme less than the cavity photon leakage, which is due to the large detuning condition. However, numerically speaking, the TQD scheme for generating two-atom 3D entanglement is very robust against decoherence induced by atomic spontaneous emissions and the cavity photon leakage, because the final fidelity is still near $0.97$ even when $\kappa=0.01g$ and $\gamma=0.05g$.

\section{Conclusion}\label{d}
In conclusion, we have proposed an experimentally feasible TQD scheme for rapidly generating two-atom 3D entanglement with one step. Not only is the scheme fast, but also the driving pulses are complete superimpositions of Gaussian pulses, which enhances the experimental feasibility greatly. Besides, adequate numerical simulations show that the TQD scheme for rapidly generating two-atom 3D entanglement is robust against deviations of the control parameters and decoherence induced by atomic spontaneous emissions and the cavity photon leakage. If adopting $^{87}$Rb~\cite{STO2003} and a set of cavity QED predicted parameters $g=2\pi\times750$MHz, $\gamma=2\pi\times3.5$MHz and $\kappa=2\pi\times2.62$MHz~\cite{STK2005} to implement the scheme, the target 3D entanglement can be obtained with $F(t_f)=0.991$, which indicates the TQD scheme for generating two-atom 3D entangled states is feasible in experiment.
\begin{center}
{\bf{ACKNOWLEDGMENT}}
\end{center}
This work was supported by the National Natural Science Foundation of China under Grants No. 11464046.

\end{document}